# Random Variables aren't Random

Paul W. Vos

February 9, 2025

**Abstract**

This paper examines the foundational concept of random variables in probability theory and statistical inference, demonstrating that their mathematical definition requires no reference to randomization or hypothetical repeated sampling. We show how measure-theoretic probability provides a framework for modeling populations through distributions, leading to three key contributions. First, we establish that random variables, properly understood as measurable functions, can be fully characterized without appealing to infinite hypothetical samples. Second, we demonstrate how this perspective enables statistical inference through logical rather than probabilistic reasoning, extending the *reductio ad absurdum* argument from deductive to inductive inference. Third, we show how this framework naturally leads to information-based assessment of statistical procedures, replacing traditional inference metrics that emphasize bias and variance with information-based approaches that better describe the families of distributions used in parametric inference. This reformulation addresses long-standing debates in statistical inference while providing a more coherent theoretical foundation. Our approach offers an alternative to traditional frequentist inference that maintains mathematical rigor while avoiding the philosophical complications inherent in repeated sampling interpretations.

## 1 Introduction

Statistical inference aims to draw conclusions about populations from observed data, yet its foundational concepts are often obscured by an unnecessary focus on randomization and hypothetical repeated sampling. This focus has led to persistent confusion and controversy in the interpretation of basic statistical concepts, even among experienced researchers. A striking example appears in discussions of p-values, where emphasis on random phenomena and hypothetical datasets clouds the simpler mathematical reality: a p-value is fundamentally a measure of location in a sampling distribution. The confusion surrounding p-values illustrates a broader issue in statistical inference – the tendency to explain mathematical concepts through mental constructs involving infinite hypothetical samples rather than recognizing them as well-defined mathematical objects.

The confusion arising from emphasis on randomization is illustrated by Gelman & Loken (2014) who begin their article stating that "Researchers typically express the confidence in their data in terms of p-value: the probability that a perceived result is actually the result of random variation." This characterization obscures the mathematical nature of p-values by casting them as descriptions of random phenomena which is but one potential use, rather than what they are: measures of where the observed sample lies in a specifically defined sampling distribution.

The authors further assert that "p-values are based on what would have happened under other possible data sets." This interpretation moves the discussion from mathematics to what Penrose (2007) calls the mental world — a realm of imagination and hypothetical scenarios. Such a transition inevitably leads to confusion. A p-value is purely mathematical: it represents the percentile of the observed sample in the distribution of all possible samples under the null hypothesis model. This distribution and the resulting p-value are part of mathematics. While proper physical randomization of the observed sample is crucial for connecting the p-value to the actual population, this requires only the single randomization that produced our data.

While Gelman & Loken (2014) make valid points about potential misuse of statistical methods, their conclusion that "the justification for p-values lies in what would have happened across multiple data sets" reveals how deeply embedded the repeated sampling perspective has become. The problems they identify stem not from p-values themselves, which are well-defined mathematical quantities, but from interpreting these mathematical objects through the lens of hypothetical repeated sampling. When we maintain focus on the mathematical structure – the distribution specified by the null hypothesis and the location of our observed sample within its sampling distribution – the meaning of p-values becomes clearer and their proper use more evident.

The remainder of this paper develops these ideas systematically. Section 2 introduces distributions on finite sets, called simple distributions, and shows how these extend naturally to both countable and uncountable sets without requiring the concept of randomness. Section 3 employs Penrose's distinction between mathematical, physical, and mental worlds to clarify how single-instance physical randomization validates our use of mathematical sampling distributions. Section 4 shows how measure theory provides precise tools for describing data through percentiles and tail areas, laying groundwork for the logical approach to inference developed in Section 5. There, we show how the *reductio ad absurdum* argument from deductive logic extends naturally to statistical inference, establishing what we call the Fisher-Information-Logic (FIL) approach. Section 6 examines traditional approaches to inference based on bias and variance, revealing their limitations by focusing on additional structures of the support set rather than the probabilities assigned to the support. Section 7 describes the information-based framework that better aligns with the mathematical nature of distributions while providing practical tools for inference.

# 2 Mathematical Framework of Distributions

## 2.1 Simple Distributions

Simple distributions provide the mathematical foundation for describing finite populations, forming the basis for more abstract probability concepts. We begin with this concrete case because it captures the essential features of distributions while maintaining clear connections to observable populations.

Let $f^1, f^2, \ldots, f^K$ be positive integers that sum to $N$, representing frequencies in a population of size $N$. For example, these might represent counts of different blood types in a hospital's patient database or the number of students receiving each possible grade in a class. Let $\mathcal{X}_K$ be a set with $K$ distinct elements. For example, $\mathcal{X}_K$ might consist of labels for the different blood types or letter grades. We define an $(N, K)$-*simple frequency distribution* as:

$$\{(x_1, f^1), (x_2, f^2), \ldots, (x_K, f^K)\}$$

where each $x_k \in \mathcal{X}_K$ and each $f^k$ represents its frequency in the population.

A *simple distribution* normalizes these frequencies by the population size $N$. Formally, it is a function $m_{(N,K)} : \mathcal{X}_K \to (0, 1]$ satisfying:

- $\sum m_{(N,K)}(x) = 1$
- $N m_{(N,K)}(x) \in \mathbb{Z}^+$ for all $x \in \mathcal{X}_K$

The distribution is degenerate if $m_{(N,K)}(x) = 1$ in which case $K = 1$.

The support $\mathcal{X}_K$ is an abstract set. To emphasize that there may be additional structure on this set, $\mathcal{X}_K$ is called the *label space* and the structure will range from no structure to the algebraic properties of the reals when $\mathcal{X}_K \subset \mathbb{R}$. When the labels have meaningful numeric values, the simple distribution will be approximated by a continuous distribution described in Section 2.3.

Each simple distribution corresponds to a unique multi-set (or bag) $\lfloor m_{(N,K)} \rfloor$ containing $N$ elements, where each value $x_k$ appears exactly $f^k$ times. Formally, this multi-set can be represented as a set of $N$ ordered pairs where the first component takes values in $\mathcal{X}_K$ and the second component ranges over $\{1, \ldots, N\}$ so that each order pair is unique. For brevity, we write $\lfloor m \rfloor$ when $N$ and $K$ are clear from context.

The multi-set shows the connection with the more common notation $X$ for a distribution. For simple distributions, $X = \Pi_1 \lfloor m \rfloor$ where $\Pi_1$ is the projection onto the first component. The proportion corresponding to the value $x$ is obtained from the counting measure of its pre-image

$$m(x) = |X^{-1}(x)|/N.$$

To avoid thinking about $X$ as describing a random process it is helpful to have a visualization. The graphical representation of a simple distribution depends on the structure of the label space $\mathcal{X}_K$:

- *For ordered* $\mathcal{X}_K$ (e.g., course grades A, B, C, D, F): Plot points on the horizontal axis with uniform spacing, constructing rectangles centered at each $x$ with area $m_{(N,K)}(x)$.

- *For unordered* $\mathcal{X}_K$ (e.g., blood types): The visualization uses similar rectangles, but their horizontal arrangement carries no meaningful information.

- *For* $\mathcal{X}_K \subset \mathbb{R}$ (e.g., height measurements): The horizontal spacing reflects the numerical values in $\mathcal{X}_K$, with rectangle heights chosen to achieve areas of $m_{(N,K)}(x)$.

The label space $\mathcal{X}_K$ and corresponding proportions $m(x)$ shown in these visualizations are the defining features of general distributions which we consider next.

## 2.2 Discrete Distributions

Simple distributions model finite populations where $N$, while typically large, remains unspecified. Discrete distributions generalize this concept by removing the dependence on $N$.

Formally, a discrete distribution on $\mathcal{X}_K$ is a function $m_K : \mathcal{X}_K \to (0, 1]$ satisfying $\sum_{\mathcal{X}_K} m_K(x) = 1$. The space of discrete distributions on $\mathcal{X}_K$ encompasses all $(N, K)$-simple distributions for $N \geq K$. Each non-degenerate discrete distribution corresponds to a point in the open simplex $\Delta_{(K-1)} \subset \mathbb{R}^K$.

When $\mathcal{X}_K$ lacks structure beyond that of an ordering, there is a natural bijection between the space of discrete distributions on $\mathcal{X}_K$ and the simplex $\Delta_{(K-1)}$. This geometric perspective reveals that families of distributions on $\mathcal{X}_K$ typically form smooth submanifolds in $\Delta_{(K-1)}$, a structure not available when restricted to simple distributions.

The notation $X_K$ for the distribution $m_K$ emphasizes the label space. Formally, $\Pr(X_K = x) = m_K(x)$ where $\Pr$ indicates measure-theoretic probability; that is, a generalization of a proportion rather than a number that describes a random phenomenon.

The next generalization is to allow $K$ to be arbitrarily large so that the support, $\mathcal{X}_\mathbb{N}$, is countably infinite. That is, $|\mathcal{X}_\mathbb{N}| = |\mathbb{N}|$ where $\mathbb{N} = \{1, 2, 3, \ldots\}$.

## 2.3 Continuous Distributions

Continuous distributions arise naturally when modeling measured values with specified precision or grouped data where $K$ is large but imprecise. These distributions often serve as approximations to simple distributions, particularly when dealing with physical measurements.

For an open interval $\mathcal{X} \subset \mathbb{R}$, a random variable $X$ with probability density function (pdf) $m$ is visualized as a curve over $\mathcal{X}$ with total area 1. Areas under portions of this curve approximate the rectangular areas of corresponding simple distributions. This approximation becomes increasingly accurate as the measurement precision increases and the population size grows.

As with simple and discrete distributions, the notation $X$ for $m$ emphasizes the label space which in this case has the advantage of representing the algebraic structure inherited from $\mathbb{R}$. Formally, $\Pr(X \in A) = \int_A m(x)dx$ for measurable set $A$. The support $\mathcal{X}$ for continuous distributions is uncountable; countability is achieved by requiring a $\sigma$-algebra on $\mathcal{X}$. It is common practice to use $X$ to denote both continuous and discrete distributions. We will do the same when the context indicates the cardinality of the support.

## 3 Randomization and the Three Worlds

Penrose's distinction between the physical, mental, and mathematical worlds provides a framework for understanding the role of randomization in statistical inference. Drawing from his *The Road to Reality* (Penrose, 2007), these worlds can be characterized as follows:

- The *physical world* contains observable phenomena and tangible reality, including the actual process of random selection through mechanisms like shuffling cards or rolling dice

- The *mental world* encompasses human consciousness, understanding, and imagination, including our intuitive conceptualization of probability and randomness

- The *mathematical world* consists of absolute, objective truths that exist independently of human thought or physical reality, including the formal structures of measure theory and probability measures

Importantly, randomization exists only in the physical world as an actual process. While the mental world contains our understanding and intuition about randomness, these are distinct from physical randomization itself. Distributions are part of mathematics that can serve as models for random phenomena in the physical world but they are also models for real-world populations.

There is a direct connection between the sample $y$ of size $n$ in the physical world and a simple distribution in the mathematical world. Let $X_{(N,K)}$ be the distribution that provides an exact model for the real world population and $\lfloor X \rfloor$ be its bag of $N$ elements. Let $\lfloor X \rfloor^{(n)}$ be the bag of all subsets of $\lfloor X \rfloor$ of size $n$. The ordered pairs in $\lfloor X \rfloor^{(n)}$ consist of $n$-tuples where the second component ranges over all $\frac{N!}{(N-n)!}$ $n$-tuples from $\{1, 2, \ldots, N\}$. The distribution $Y_{(N',K')}$, in particular the values for $N'$ and $K'$, for the bag $\lfloor X \rfloor^{(n)}$ is obtained using combinatorics.

If $y$ was obtained using a simple random sample, that is, if it was chosen in such a way that every sample of size $n$ had an equal chance of being selected, then $Y_{(N',K')}$ is the exact distribution for $y$. For sampling plans other than simple random samples, $\lfloor X \rfloor^{(n)}$ is replaced with a different subset of the powerset of $\lfloor X \rfloor$. The proportions obtained by combinatorics are what connect the simple distributions to the physical and mental worlds. This connection is fundamental to understanding statistical inference in terms of distributions of data rather than models of random phenomena, i.e., random variables.

Measure theory extends the idea of a proportion of a finite set to a probability that describes an infinite set. This extension does not need to occur in the mental or physical world and attempts to do so using hypothetical repeated sampling from a real-world population is ripe for confusion.

# 4 Probability to Describe Observed Data

## 4.1 Proportions and Percentiles

For a simple distribution $X_{(N,K)}$, proportions arise naturally from relative frequencies. The proportion of any value $x$ is defined as $f(x)/N$, where $f(x)$ represents its frequency. These proportions form the basis for understanding more abstract probability concepts.

When the support set $\mathcal{X}$ carries an ordering, we can define the percentile of a value $x$ as $100 \sum_{x' \leq x} f(x')/N$. This percentile characterizes the location of $x$ within the distribution $X$. An equivalent and often more useful characterization comes through tail areas:

- Left tail area: $TA_L(x) = \sum_{x' \leq x} f(x')/N$
- Right tail area: $TA_R(x) = \sum_{x \leq x'} f(x')/N$
- Tail area: $TA(x) = 2 \min\{TA_L(x), TA_R(x)\}$

For continuous distributions the sums are replaced with integrals.

The concept of extreme values in a distribution warrants careful consideration. Values with small tail areas are termed extreme, and while these might be called unlikely or surprising, such characterizations describe the value's location within the distribution rather than any inherent probability of the specific observation. A better synonym for 'extreme' is 'rare' indicating a property shared by a relatively few members of a population. Consider human height: meeting someone exceptionally tall represents an extreme or rare event because of where that height sits in the population distribution, not because of any inherent improbability.

This distinction becomes particularly clear in games of chance, such as 5-card poker. In a well-shuffled deck, each specific 5-card hand occurs with equal probability $1/\binom{52}{5}$. However, the game of poker requires an imposed ordering on equivalence classes of hands. Consider two hands: $H_1 = \{2\heartsuit, 3\heartsuit, 4\heartsuit, 5\heartsuit, 7\clubsuit\}$ and $H_2 = \{2\spadesuit, 3\spadesuit, 4\spadesuit, 5\spadesuit, 6\spadesuit\}$. While these hands have identical probabilities of being dealt, $H_2$ lies further in the tail of the distribution of hand rankings, as significantly fewer hands beat it. A hand this far into the ordering is a rare hand.

## 4.2 Fisher's Infinite Population

R.A. Fisher did not adhere to the view of probability as describing hypothetical repeated samples, instead, as Spiegelhalter (2024) notes, "Fisher suggested thinking of a unique data set as a sample from a hypothetical infinite population, but this seems to be more of a thought experiment than an objective reality." However, Penrose's distinction between mathematical and mental worlds provides an objective reality to Fisher's framework.

To understand Fisher's approach, we begin with a simple distribution $X_{(N,K)}$ that exactly models a finite population. The observed data $y$ represents a point in $Y_{(N,K)}$, the sampling distribution of $X_{(N,K)}$. When we approximate $X_{(N,K)}$ with a continuous distribution $X$, the corresponding sampling distribution $Y$ approximates $Y_{(N,K)}$. Both $X$ and $Y$ will have infinite support.

The infinite sampling distribution exists in the mathematical world rather than the mental world. It is not a hypothetical construct requiring repeated sampling or imagination, but rather a mathematical object with the same objective reality as any other mathematical structure. Fisher's infinite population, therefore, describes a precise mathematical framework for understanding sampling distributions, not a mental exercise in hypothetical repetition.

# 5 Logic of Statistical Inference

## 5.1 Deduction and Induction

We use the following example to illustrate the distinction between deductive and inductive reasoning in statistical inference.

**Example 1.** Two Lotteries

Consider a scenario involving an extraterrestrial civilization comprised of two distinct nations, both of which have discovered Earth. While these nations coexist peacefully, they hold divergent views regarding humanity, with Nation A representing an existential threat to Earth.

Earth's intelligence services have intercepted communications revealing that both nations operate similar lottery systems. Each nation conducts a pick-4 lottery where four numbers are drawn from separate but identical bins, differing only in the numerical range of their balls. Nation A's lottery uses balls numbered 0 through 7, while Nation B employs balls numbered 0 through 9. Intelligence reports indicate that a spacecraft from one of these nations is approaching Earth, commanded by a captain known for wagering on the sum of their national lottery numbers. The next communication is expected to contain this sum.

Earth's scientific community has determined that if the received sum exceeds 28, this conclusively identifies Nation B as the approaching civilization. This conclusion follows from a *reductio ad absurdum* argument: assume the sum originates from Nation A's lottery, where each draw must be between 0 and 7. The maximum possible sum would be 28 (achieved when drawing 7 four times).

Therefore, any sum exceeding 28 contradicts the assumption that it came from Nation A, leaving Nation B as the only possible source.

This argument can be expressed in terms of distributions as follows. Let $\mathcal{S}_k = \{k' \in \mathbb{Z} : 0 \leq k' \leq k\}$ be the set of non-negative integers up through $k$. Let $S$ be the distribution for the observed sum $s_{obs}$. The premise, known as the null hypothesis in statistical terminology, is $H_\circ : S = S_A$ where $S_A$ is the distribution of sums from lottery A whose support is $\mathcal{S}_A = \mathcal{S}_{28}$. The steps of the argument for a sum of 29 are as follows:

1. Begin with $H_\circ$

2. $H_\circ \implies s_{obs} \in \mathcal{S}_{28}$; $s_{obs} = 29$

3. $s_{obs} \notin \mathcal{S}_{28}$. A contradiction. $\therefore \neg H_\circ$

This argument provides proof that $H_\circ$ is false; i.e. $S \neq S_A$. Clearly, this argument goes through whenever $s_{obs} \notin \mathcal{S}_{28}$.

When the observed sum is 28 or less, reasoning shifts from deduction to induction, revealing fundamental divisions within the statistical community. The repeated-sampling frequentists either conclude that no inference is possible or resort to multiverse arguments to maintain their philosophical framework. The Bayesian school insists that prior probabilities must be assigned to the two possibilities, yet this leads to considerable debate regarding both the numerical values of these priors and their philosophical interpretation. Good (1971), himself a Bayesian, wryly observes that there are at least "46,656 varieties of Bayesians".

## 5.2 Induction and Logic

A third approach emerges from those who recognize that the deductive argument exists purely within mathematics and can be extended using mathematical principles, providing an objective framework for analysis. This mathematical extension is motivated by the work of Fisher (one reference being pages 42-44 of Fisher (1959)).

The deductive argument only required $\mathcal{S}_{28}$, the support of the distribution specified by $H_\circ$. The inductive argument requires $S_A$ the distribution specified by $H_\circ$ along with another distribution that describes lottery B conditioned on the event $s_{obs} \in \mathcal{S}_{28}$. This conditioning reflects the fact that we use a deductive argument when $s_{obs} \notin \mathcal{S}_{28}$. For simplicity of notation we use $S_B$ for this conditional distribution.

This extension introduces an auxiliary hypothesis $H_{aux}$: "the observed sum is not exceptionally rare." This auxiliary hypothesis formalizes a principle common to human experience - we generally assume that exceptionally rare events do not occur, as evidenced by our willingness to engage in activities like air travel despite infinitesimal but non-zero risks.

While the concept of "rare" might initially seem subjective, it can be quantified objectively using tail areas of a distribution. The tail area - specifically, the proportions associated with extreme values - provides a mathematical framework for defining rarity. For instance, if we designate a sum of 28 as rare but 27 as not rare for the distribution $S_A$, this corresponds to labeling the uppermost 0.0002 of possible sums for lottery A as rare. More formally, we can express $H_{aux}$ as the statement that $s_{obs}$ falls below the 99.98th percentile using the natural ordering provided by the numerical values of the sum.

This formalization allows us to construct a *reductio ad absurdum* argument applicable for induction:

1. Begin with the conjunction $H_\circ \land H_{aux}$

2. This conjunction implies $s_{obs} \in \mathcal{S}_{27}$; $s_{obs} = 28$

3. $s_{obs} \notin \mathcal{S}_{27}$. A contradiction. $\therefore \neg(H_\circ \land H_{aux}) \equiv \neg H_\circ \lor \neg H_{aux}$

This argument structure is particularly noteworthy because both hypotheses play essential roles: $H_{aux}$ establishes both the ordering and the numerical threshold for extreme values, while $H_\circ$ determines the proportions assigned to these values. The conclusion takes the form of a logical disjunction that Fisher (1960) described as:

> Either the hypothesis is not true, or an exceptionally rare outcome has occurred.

The definition of rare events through tail areas can be calibrated to different levels of evidence. For example, if we consider a sum of 24 as extreme but 23 as not, this corresponds to a tail area of 0.0171 for lottery A.

This framework provides a bridge between deductive and inductive reasoning. Rather than requiring the conceptual machinery of hypothetical repeated sampling or subjective prior probabilities, it extends classical deductive logic to accommodate uncertainty in a mathematically rigorous way. The approach maintains objectivity while acknowledging the inherent limitations in our ability to draw absolute conclusions from empirical data.

### 5.3 Test Statistics

The transition from deductive to inductive reasoning in statistical inference highlights two fundamental requirements. First, we must assume that the distribution of observed data matches that of the underlying population - an assumption validated through proper randomization. Second, we require a method for ordering possible outcomes to evaluate their extremity. This ordering represents an important choice in statistical analysis.

The mathematical tool for imposing this order is the test statistic. Formally, a test statistic for hypothesis $H_\circ$ is a real-valued function $T$ defined on the sample space. For each value $t$ in the image of $T$, the pre-image $T^{-1}(t)$ forms a subset of the sample space. As $t$ ranges over all possible values, these pre-images partition the sample space into subsets ordered by $t$. When working with continuous distributions, we require $T$ to be measurable to ensure compatibility with the probability structure.

There are many choices for the test statistic and different statistics provide different orderings of the sample space. In our Two Lotteries example, we seek a test statistic that effectively distinguishes between distributions $S_A$ and $S_B$.

The Neyman-Pearson lemma provides theoretical guidance for this choice, showing that the likelihood ratio test achieves optimal power when comparing two simple hypotheses, as in our lottery example. Interestingly, for sums in $\{8, 9, \ldots, 27, 28\}$, the ordering induced by the likelihood ratio matches that of the numerical values themselves. However, the likelihood ratio remains constant across $\{0, 1, \ldots, 6, 7\}$, revealing a subtle distinction between natural and likelihood ratio ordering of outcomes.

Most statistical applications differ from our example in two important ways. First, the exact values of $N$ and $K$ characterizing the simple distribution are typically unknown, necessitating the use of approximating distributions $X$. For discrete $X$, the sample space $\mathcal{X}$ becomes a countable set with proportions replaced by real numbers in the unit interval. For continuous $X$, $\mathcal{X}$ becomes a measurable space with proportions replaced by measure-theoretic probability of measurable sets.

Second, rather than choosing between two specific distributions, we consider a smooth continuum of possible models. Nevertheless, the modified *reductio ad absurdum* argument extends naturally to this setting after appropriate adjustments for measurable sets. The likelihood continues to play a central role, with the smooth structure of the model space allowing us to analyze how likelihood functions vary across models - essentially a local version of the likelihood ratio.

The information content of a test statistic provides a mathematical framework for describing its induced ordering of the sample space. This connection between test statistics and information theory is described in Section 7.

## 6 Inference and Hypothetical Repeated Sampling

The traditional frequentist approach to statistical inference is motivated by a conceptual framework of hypothetical repeated sampling. This section presents this framework and shows how it naturally leads to concepts like bias and variance for evaluating statistical procedures. We purposefully maintain the repeated sampling perspective and 'random variable' notation $X$ here to demonstrate how it shapes statistical thinking, before introducing an alternative approach in Section 7.

The population we study is finite and can be exactly described by a simple distribution, which we denote by $m_{pop}$ or by $X_{pop}$ to emphasize the label space. To develop statistical procedures, we introduce a family of distributions $M$ that typically does not contain simple distributions, but does contain a distribution $X_\star$ that best approximates $X_{pop}$ and we assume it is a suitable approximation so that $X_\star$ is considered the distribution from which the sample was taken.

A parameterization is a function $\theta : M \to \Theta \subset \mathbb{R}^d$ and the goal of inference is to use the sample $y = (x_1, x_2, \ldots, x_n)$ to obtain a value for the parameter, called an estimate, that will be close to $\theta_\star = \theta(X_\star)$. The value of the estimate will depend on the sample and a desirable property is that, were the process repeated, the average of the estimates would be close to $\theta_\star$. Conceptually, this requires hypothetically sampling from the real-world population and letting the number of such samples tend to infinity before the average equals $\theta_\star$.

Mathematically, an *estimate* is the value of a measurable function $t : \mathcal{Y} \to \Theta$. The distribution that is obtained when $t$ is applied to $Y$ is an *estimator*, $\hat{\theta} = t(Y)$. Unbiasedness is described using the expectation operator, which for continuous distributions, is defined as $Eh(Y) = \int h(y) m^{\mathcal{Y}}(y) dy$ where $m^{\mathcal{Y}}$ is the density function for the distribution $Y$ obtained from $m$. This operator is defined for each distribution in $M$ so that a subscript on the operator will indicate a specific distribution. The conceptual construction of unbiasedness is expressed mathematically as $E_\star \hat{\theta} = \theta_\star$ where $E_\star$ is the expectation defined using $m = m_\star$. Our notation does not distinguish between an estimate $\hat{\theta} = t(y)$ and an estimator $\hat{\theta} = t(Y)$ but will be clear from the context. In particular, expectation operates on estimators.

The bias is the difference between this expectation and $\theta_\star$,

$$Bias_\star(\hat{\theta}) = E_\star \hat{\theta} - \theta_\star.$$

While the conceptual construction focuses on the distribution $m_\star$, mathematically bias is defined for all distributions in $M$. Since $\theta_\star$ is unknown, we want estimators that are unbiased for all values of the parameter (i.e., distributions in $M$). The estimator $\hat{\theta}$ is unbiased for $\theta$ is its bias vanishes for all values of $\theta$; that is, $Bias(\hat{\theta})$ is the zero *function* on $\Theta$.

Another important property is that the value of the estimate is, in some sense, close to $\theta_\star$. Recognizing that the estimate cannot be close for all $y \in \mathcal{Y}$, the estimator is described in terms of its average distance from $\theta_\star$ and, as with bias, a mental conceptualization of this involves hypothetical repeated samples from the physical population.

Mathematically, the distance is defined using a non-negative function $d$ defined on $\Theta \times \Theta$ so that for a fixed $y$ the distance between the estimate and $\theta_\star$ is $d(\hat{\theta}, \theta_\star)$. The average distance is found using the estimator and the expectation operator $E_\star d(\hat{\theta}, \theta_\star)$. Since $E$ and $d$ are defined for all $\theta$, $Ed(\hat{\theta}, \theta)$ is a function on $\Theta$.

The most common choice for $d$ is square error, $d(\hat{\theta}, \theta) = (\hat{\theta} - \theta)^2$, and this average distance is called mean square error

$$MSE(\hat{\theta}) = E(\hat{\theta} - \theta)^2.$$

MSE is a function on $\Theta$ and estimators that minimize MSE for all parameter values generally do not exist. When estimators are required to be unbiased, estimators that minimize MSE do exist for many important estimation problems. For unbiased estimators MSE equals the variance $V(\hat{\theta}) = E(\hat{\theta} - E\hat{\theta})^2$ and such estimators are called uniformly minimum variance unbiased (UMVU).

The controversy with UMVU estimators comes when there are biased estimators that have smaller MSE than the UMVU estimator at all values of the parameter. Using MSE as the estimation criterion, this indicates that the UMVU estimator should not be used, but leaves open the question of what estimator should be used as estimators minimizing MSE for all values of the parameter generally do not exist. Efron (2024) suggests the solution to this issue is to move from frequentist to Bayesian inference methods. Efron's suggestion is based on the assumption that there a problems with frequentist methods, in particular with maximum likelihood. What Efron found shocking was the existence of estimators that had smaller MSE for all values of the parameter (i.e., "always"):

> That "always" was the shocking part: two centuries of statistical theory, ANOVA, regression, multivariate analysis, etc., depended on maximum likelihood estimation. Did everything have to be rethought?

Not everything; just the role of bias and MSE. Rethinking these leads to information as a means of assessing estimators. Before describing the role of information in the next section we describe an important property shared by bias and MSE that illustrates a difficulty with these measures and the way forward.

**Units of Measurement and Invariance**

The fundamental requirement that statistical inference should not depend on units of measurement leads us to examine the behavior of our estimation criteria under different transformations. While bias and MSE exhibit invariance under linear transformations (allowing, for instance, conversion between kilometers and miles), they fail under more complex transformations that are equally valid representations of the same physical quantity.

Consider, for example, the measurement of fuel efficiency in automobiles. In the US, the units of measure are mile per gallon (mpg) while in the UK liters to drive 100 km (L/100 km) is used. The issue is not metric versus English units so we simplify and consider two studies of the same data that use the same family of models. In one, the data are presented in km/L while in the other the units are L/km so the units require a reciprocal transformation.

Suppose both studies used unbiasedness to choose an estimator. If the study using units km/L finds an unbiased estimator, that estimator will not be unbiased in the reciprocal units. This means that the analysis now depends on the units of measurement. If both studies use MSE and the study using km/L units finds one estimator having smaller MSE than another, this relationship need not hold using the L/km units. Again, the analysis will depend on the units chosen to measure the data.

This example illustrates deficiencies of bias and MSE to assess estimators but is also suggests a solution. Namely, ignore the structure of the label space and use the probability or probability density to assess estimators. In terms of the visualization of simple distributions using rectangles placed on the horizontal axis, we should ignore the location and other structure of this axis and focus on the height of the rectangles. The base of the rectangles serve only as an index set to compare distributions in terms of their corresponding heights.

# 7 Inference and Information

Section 6 presented the traditional frequentist framework where inference rests on hypothetical repeated sampling, leading naturally to concepts like bias and variance for evaluating statistical procedures. This section develops an alternative approach that aligns with our view of random variables as mathematical structures. Rather than considering the behavior of procedures across hypothetical samples, we fix the observed sample and examine its relationship to possible models. This shift in perspective leads to two key developments: a logical framework for inference through generalized estimation, and an information-theoretic assessment of estimators that replaces bias and variance. Technical details of generalized estimation are given in Vos (2022) and Vos & Wu (2024). See Vos & Holbert (2022) for additional discussion regarding the adequacy of a single random sample for inference.

## 7.1 From Repeated Sampling to Logical Inference

Traditional inference metrics like bias and variance arise naturally when considering sampling distributions, but they reflect properties of the label space rather than the mathematical structure required to compare distributions. The Fisher-Information-Logic (FIL) approach shifts focus to the relative frequencies (or probabilities) assigned to points in the sample space, leading to metrics that better align with the mathematical nature of distributions.

Consider how bias assessment typically works: we imagine repeatedly drawing samples from some true distribution, computing our estimate each time, and comparing the average estimate to the true parameter value. This framework requires us to reason about samples we never observe. The FIL approach instead starts with our actual observed sample $y_{obs}$ and examines its relationship to every model in a family of distributions $M$. A key point here is that the sampling distribution for each distribution in $M$ is part of mathematics requiring no sampling. The justification for comparing $y_{obs}$ to each sampling distribution depends only on the single sampling mechanism used in the physical world that resulted in $y_{obs}$.

For a smooth family of distributions $M$ with sampling distribution support $\mathcal{Y}$, we introduce generalized estimators that map $\mathcal{Y} \times M$ to $\mathbb{R}$. A generalized estimator $g$ evaluated at our observed sample $y_{obs}$ provides a function $g_{obs} = g(y_{obs}, \cdot)$ that orders all models in $M$ according to their consistency with $y_{obs}$. Rather than asking about the long-run behavior of $g$ across hypothetical samples, we examine how effectively it discriminates between different possible models given our observed data.

## 7.2 Extending Logical Arguments to Statistical Inference

The FIL approach extends the modified *reductio ad absurdum* argument from Section 5.2 to continuous families of distributions. For each model $m \in M$, we consider the conjunction of two hypotheses:

- $H$: The model $m$ best approximates the true simple distribution
- $H_{aux}$: The observed sample $y_{obs}$ is not rare under model $m$

We formalize "rare" through tail areas, which naturally extend from simple distributions to general distributions via measure theory:

$$TA_L(y_{obs}, m) = \int_{y:g(y,m) \leq g_{obs}(m)} m^{\mathcal{Y}} dy$$

where $m^{\mathcal{Y}}$ represents the sampling distribution under $m$. The right tail area $TA_R$ uses $\geq$ instead of $\leq$, and we set $TA(y_{obs}, m) = 2\min(TA_L, TA_R)$. For significance level $\alpha$, we formalize $H_{aux}$ as $TA(y_{obs}, m) > \alpha$.

This framework partitions the model space into two sets: $M_\alpha^{\complement}$ containing all distributions where the *reductio ad absurdum* argument successfully reaches a contradiction (the observed data are rare), and $M_\alpha$ containing the distributions where the argument fails to reach a contradiction::

$$\begin{aligned} M_\alpha^{\complement} &= \{m \in M : TA(y_{obs}, m) \leq \alpha\} \\ M_\alpha &= \left\{m \in M : m \notin M_\alpha^{\complement}\right\} \end{aligned}$$

We cannot know with certainty which of these sets contains $m_\star$, the distribution in $M$ that best approximates the true simple distribution $X_{pop}$. However, if $m_\star \in M_\alpha^{\complement}$ then $y_{obs}$ is rare. We have Fisher's logical disjunction: either the true distribution is in $M_\alpha$ or the sample we observed is rare. The set $M_\alpha$ forms a $(1-\alpha)100\%$ confidence region, with its image under a parameterization $\theta$ giving a subset $\Theta_\alpha \subset \mathbb{R}^d$. When $d = 1$, $\Theta_\alpha$ often forms an interval, in which case we call it a confidence interval.

## 7.3 Information Theory and Statistical Evidence

The FIL approach centers on comparing distributions rather than focusing on a single hypothetical true distribution. This shift naturally leads to information-theoretic concepts for measuring the strength of statistical evidence. While tail areas provide evidence against specific distributions through the *reductio ad absurdum* argument, the effectiveness of a generalized estimator is measured through its information content $\Lambda(g)$. This connection between statistical evidence and information mirrors fundamental concepts in information theory: the Kullback-Leibler (KL) divergence (also called KL information) and entropy.

Like our treatment of generalized estimators, these information-theoretic measures depend only on probability assignments, not on the structure of the label space. For distributions sharing support $\mathcal{X}$, the KL divergence:

$$KL(m_1, m_2) = \sum_{x \in \mathcal{X}} m_1(x) \log\left(\frac{m_1(x)}{m_2(x)}\right)$$

measures the information lost when $m_2$ is used to approximate $m_1$. Similarly, entropy:

$$ENT(m) = -\sum_{x \in \mathcal{X}} m(x) \log m(x)$$

quantifies the information content of a distribution using only its probability structure. These equations illustrate how $\mathcal{X}$ serves purely as an index set, consistent with our emphasis on probability assignments over label space structure.

The Fisher information and the information utilized by generalized estimators, $\Lambda(g)$, share this independence from label space structure, requiring only measurability for continuous distributions. However, they differ from KL divergence and entropy in a crucial way: while KL and entropy can be computed for individual distributions or pairs of distributions, Fisher information and $\Lambda(g)$ require a smooth family of distributions. This reflects their roles in statistical inference, where they describe properties of distribution families and functions defined on these families rather than isolated distributions.

When distributions have different supports, the KL divergence becomes infinite — precisely the cases where deductive rather than inductive reasoning applies, as illustrated in our Two Lotteries example. This connection between information theory and logical inference reinforces how the FIL approach provides a unified framework for statistical reasoning that maintains mathematical rigor while avoiding the conceptual burden of hypothetical repeated sampling.

## 8 Discussion

This paper has developed a framework for understanding statistical inference that emphasizes the mathematical nature of distributions while carefully distinguishing between mathematical, physical, and mental worlds. This distinction proves particularly valuable when examining fundamental statistical concepts and terminology that often conflate these domains. Three key areas illustrate both the importance and broader implications of this perspective.

First, statistical terminology frequently introduces mental world connotations that can obscure rather than clarify mathematical concepts. The term "random variable" exemplifies this problem - while mathematically defined as a measurable function, the word "random" suggests a mental world construct involving chance and unpredictability. Similar issues arise with terms like "information," "likelihood," and "confidence." These terms carry intuitive meanings that may mislead practitioners about their precise mathematical definitions. The solution is not to eliminate such terminology - these terms are deeply embedded in statistical practice - but rather to explicitly recognize and address the potential confusion they may cause.

Second, the concept of information in statistics requires particular care. While information theory provides precise mathematical definitions through concepts like entropy and KL divergence, these capture only specific aspects of how information is understood more broadly.

The phrase "amount of information" might suggest information is a quantity like mass or volume, but this analogy breaks down upon closer examination. Information does not have units, and the relationship between information and variance illustrates this subtlety.

Consider an unbiased estimator $\hat{\theta}$ where the information $\Lambda(\hat{\theta})$ equals the reciprocal of its variance. While these quantities are numerically reciprocal, they are conceptually distinct: $\Lambda(\hat{\theta})$ measures how rapidly probability assignments change when considering different models, while variance is defined for an isolated distribution and quantifies the spread using the algebraic properties of the label space. The parameter $\theta$ plays fundamentally different roles in each case - for information, it serves as a coordinate on a smooth manifold (and thus has no units), while for variance, it inherits the units and structure of the label space. See Vos (2024) for problems when inference depends on the choice of parameterization.

Third, our framework provides new insights into point estimation. When working with continuous exponential families, we can leverage the bijection between canonical statistics and expectation parameters to view points in the support $\mathcal{X}$ as either canonical statistics or expectation parameters. In either case, $\mathcal{X}$ becomes a subset of $\mathbb{R}^d$, which we denote as $\mathcal{X}_{\mathbb{R}}$. More fundamentally, using the parameterization allows us to take $\mathcal{X} = M$, viewing a point estimate as a distribution in the model space $M$ rather than as a real number labeling this distribution. While $M$ lacks the algebraic structure of the reals, concepts like mean and variance can be defined through optimization, generalizing familiar ideas like least squares to other exponential families using KL divergence. See Wu & Vos (2012) and Vos & Wu (2015) for details on these distribution-valued point estimators.

These observations point to a broader principle: the importance of maintaining clear distinctions between mathematical definitions and their interpretations in the physical and mental worlds. The fact that "random variables aren't random" serves as a caution about mathematical terminology in general - terms that carry rich meaning outside mathematics may not align with their precise mathematical definitions. This misalignment becomes particularly problematic when it leads to reasoning about hypothetical scenarios (like infinite sequences of samples) rather than focusing on well-defined mathematical objects.